\documentstyle[12pt]{article}
\begin{document}

\title{Laplace transformations
of hydrodynamic type systems
in Riemann invariants: periodic sequences}
\author{{\Large Ferapontov E.V.\thanks{
    Present address:
    Fachbereich Mathematik, SFB 288,
    Technische Universit\"at Berlin,
    10623 Berlin,
     Germany,\ \ 
    \hbox{e-mail: {\tt fer@sfb288.math.tu-berlin.de}}}}\\
    Institute for Mathematical Modelling\\
    Academy of Science of Russia, Miusskaya, 4\\
    125047 Moscow, Russia\\
    e-mail: {\tt fer@landau.ac.ru}
}
\date{}
\maketitle

\newtheorem{theorem}{Theorem}

\pagestyle{plain}

\maketitle

\begin{abstract}
The conserved densities of hydrodynamic type system in Riemann invariants
satisfy a system of linear second order
partial differential equations.
For linear systems of this type Darboux introduced Laplace
transformations, generalising the classical transformations
in the scalar case. It is demonstrated that Laplace
transformations can be pulled back to the transformations of the
corresponding hydrodynamic type systems. We discuss periodic Laplace sequences of
$2\times 2$ hydrodynamic type systems with the emphasize on the simplest 
nontrivial case of  period 2.

For $3\times 3$ systems in Riemann invariants
a complete discription of closed quadruples is proposed. They turn to be
related to a special quadratic reduction of the $(2+1)$-dimensional
3-wave system which can be reduced to a triple of pairwize commuting 
Monge-Ampere equations.

In terms of the Lame and rotation coefficients Laplace transformations 
have a natural interpretation as
the symmetries of the Dirac operator, associated with the 
$(2+1)$-dimensional $n$-wave system. The 2-component Laplace transformations
can be interpreted also as the symmetries of the (2+1)-dimensional 
integrable equations of Davey-Stewartson type.

Laplace transformations of hydrodynamic type
systems originate from a canonical geometric 
correspondence between systems of conservation laws and line congruences 
in  projective space.

\end{abstract}

\section{Introduction and the main results}
Let us consider a two-component system of hydrodynamic type in
Riemann invariants
$$
\begin{array}{c}
R_t^1=\lambda^1(R)R^1_x,\\
R_t^2=\lambda^2(R)R^2_x.\\
\end{array} \eqno(1)
$$
Any system (1) possesses infinitely many conservation laws of hydrodynamic type
$\int u(R) dx$ with the conserved densities $u(R)$  satisfying 
linear equation of the second order
$$
\partial_1\partial_2u=a\partial_1u+b\partial_2u, \eqno(2)
$$
where
$$
a=\frac{\partial_2\lambda^1}{\lambda^2-\lambda^1}, ~~~
b=\frac{\partial_1\lambda^2}{\lambda^1-\lambda^2},
$$
$\partial_i=\frac{\partial}{\partial R^i}$. 
Let also $f$ be the flux, corresponding to the density $u$, that is, $f$
satisfies the equation $u_t=f_x$ or, equivalently,
$$
\partial_if=\lambda^i\partial_iu ~~~~ {\rm for ~any} ~~~ i=1, 2,
$$
which are compatible due to (2).

Applying to equation (2) the Laplace transformation
$$
U=u-\frac{\partial_1u}{b},
$$
we arrive at the similar equation with respect to $U$:
$$
\partial_1\partial_2U=A\partial_1U+B\partial_2U,
$$
where
$$
A=a-\partial _2\ln b, ~~~
B=b+\partial_1\ln A. \eqno(3)
$$
(the inverse Laplace transformation $U=u-\frac{\partial_2u}{a}$
is considered analogously).
It turns out that Laplace transformations can be pulled back to the
transformations of the corresponding hydrodynamic type systems.
Let us introduce the system
$$
\begin{array}{c}
R_t^1=\Lambda^1(R)R^1_x,\\
R_t^2=\Lambda^2(R)R^2_x,\\
\end{array} \eqno(4)
$$
where the new characteristic velocities
$\Lambda^1, \Lambda^2$ are connected with
$\lambda^1, \lambda^2$ by the formulas
$$
\begin{array}{c}
\Lambda^1=\lambda^2,  \\
\ \\
\Lambda^2=\lambda^2-\frac{\displaystyle b\partial_2\lambda^2}
{\displaystyle \partial_2b-ab}=
\lambda^2-\frac{\displaystyle 1}
{\displaystyle \frac{\partial_1\partial_2\lambda^2}
{\partial_1\lambda^2 \partial_2\lambda^2 }+
\frac{1}{\lambda^1-\lambda^2}}.
\end{array}
\eqno(5)
$$

\begin{theorem}

1. The conserved densities $U$ of system (4)
are given by the formula
$$
U=u-\frac{\partial_1u}{b},
$$
where $u$ are conserved densities of system (1).

2. The characteristic velocities $W^1, W^2$  of commuting flows of system
(4) are given by the formulas
$$
\begin{array}{c}
W^1=w^2,  \\
\ \\
W^2=w^2-\frac{\displaystyle b\partial_2w^2}{\displaystyle \partial_2b-ab},
\end{array}
$$
where $w^1, w^2$ are the characteristic velocities of
commuting flows of system (1), i.e. solutions of the linear system
(see \cite{Tsarev})
$$
\frac{\partial_2w^1}{w^2-w^1}=a, ~~~
\frac{\partial_1w^2}{w^1-w^2}=b.
$$

3. The flux $F$ of the conserved density $U$ is given by the formula
$$
F=f-\lambda^2 \frac{\partial_1u}{b},
$$
where $f$ is the flux of $u$.
\end{theorem}

The proof follows from  the identities
$$
A=\frac{\partial _2 \Lambda ^1}{\Lambda ^2-\Lambda ^1}, ~~~
B=\frac{\partial _1 \Lambda ^2}{\Lambda ^1-\Lambda ^2},
$$

$$
\frac{\partial _2 W^1}{W^2-W^1}=A, ~~~
\frac{\partial _1 W^2}{W^1-W^2}=B,
$$
and
$$
\partial_iF=\Lambda^i\partial_iU, ~~~  i=1, 2.
$$
which can be checked by a direct calculation.
It is natural to call system (4) the Laplace transformation of system
(1). Evidently Laplace transformations preserve the
"integrability": the conserved
densities and commuting flows of system (1)
are automatically transformed
into the conserved densities and commuting flows of system (4)
according to the formulas of theorem 1. In particular,
solutions of system (1),
specified by the hodograph formula (see \cite{Tsarev})
$$
w^1=x+\lambda^1t, ~~~ w^2=x+\lambda^2t,
$$
are transformed into the solutions
$$
W^1=x+\Lambda^1t, ~~~ W^2=x+\Lambda^2t,
$$
of system (4).

The inverse Laplace transformation $U=u-\frac{\partial_2u}{a}$
corresponds to the interchange of indices 1 and 2 in formulas (5).
Some further properties of Laplace transformations in the two-component case
are discussed in sect. 2. 

Following \cite{ShabYam} in sect. 5 we give the interpretation of  Laplace
transformation (5) as a discrete symmetry of the integrable equations of
Davey-Stewartson type.

Formula (5) results from the geometric construction of paper \cite{AgaFer},
relating systems of conservation laws with line congruences in the
projective space. This correspondence is briefly discussed in sect. 6.

\bigskip

Laplace transformations can be generalised to $n\times n$ systems
in Riemann invariants
$$
R_t^i=\lambda^i(R)R^i_x, \eqno(6)
$$
namely, for any pair of indices $i\ne j$ we define transformation $S_{ij}$,
mapping system (6) into the new system
$$
R_t^i=\Lambda^i(R)R^i_x, \eqno(7)
$$
with the characteristic velocities
$$
\begin{array}{c}
\Lambda^i=\lambda^j, \\
\ \\
\Lambda^j=\lambda^j-\frac{\displaystyle a_{ji}\partial_j\lambda^j}
{\displaystyle \partial_ja_{ji}-a_{ij}a_{ji}}, \\
\ \\
\Lambda^k=\frac{\displaystyle \lambda^ka_{ji}-\lambda^ja_{ki}}
{\displaystyle a_{ji}-a_{ki}}, ~~~
k\ne i, j,
\ \\
\end{array} \eqno(8)
$$
where $a_{ji}=\frac{\partial_i\lambda^j}{\lambda^i-\lambda^j}$.

\begin{theorem}

1. The conserved densities $U$ of system (7) are given by the formula
$$
U=u-\frac{\partial_iu}{a_{ji}},
$$
where $u$ are conserved densities of system (6).

2. The characteristic velocities $W^i$  of commuting flows of
   system (7) are given by the formulas
$$
\begin{array}{c}
W^i=w^j, \\
\ \\
W^j=w^j-\frac{\displaystyle a_{ji}\partial_jw^j}
{\displaystyle \partial_ja_{ji}-a_{ij}a_{ji}}, \\
\ \\
W^k=\frac{\displaystyle w^ka_{ji}-w^ja_{ki}}{\displaystyle a_{ji}-a_{ki}}, ~~~
k\ne i, j,
\ \\
\end{array}
$$
where $w^i$ are the characreristic velocities
of commuting flows of system (6),
i.e solutions of the linear system
(see \cite{Tsarev})
$$
\frac{\partial_jw^i}{w^j-w^i}=a_{ij}, ~~~ i\ne j.
$$

3. The flux $F$ of the conserved density $U$ of system (7) is
given by the formula
$$
F=f-\lambda^j \frac{\partial_iu}{a_{ji}},
$$
where $f$ is the flux of $u$.

\end{theorem}

We recall that the conserved densities $u$ of system (6)
satisfy an overdetermined system of linear second order equations
$$
\partial_i\partial_ju=a_{ij}\partial_iu+a_{ji}\partial_ju, ~~~~
i\ne j, \eqno(9)
$$
with the compatibility conditions
$$
\partial_ka_{ij}=a_{ik}a_{kj}+a_{ij}a_{jk}-a_{ij}a_{ik}, ~~~~
i\ne j\ne k\ne i,
$$
which we always assume to be satisfied. Systems (6), satisfying these
compatibility conditions,
are called semihamiltonian  and can be integrated by the
generalised hodograph transform \cite{Tsarev}. The formula
$$
U=u-\frac{\partial_iu}{a_{ji}}
$$
defines Laplace transformation $S_{ij}$ of  linear
system (9) -- see Darboux \cite{Darboux}, p.274.
As one can verify directly, $U$ satisfies the system
$$
\partial_i\partial_jU=A_{ij}\partial_iU+A_{ji}\partial_jU, ~~~ i\ne j,
$$
with the coefficients $A$ given by the formulas
$$
\begin{array}{c}
A_{ij}=a_{ij}-\partial_j\ln a_{ji}, \\
\, \\
A_{ji}=a_{ji}+\partial_i\ln A_{ij}, \\
\, \\
A_{ik}=a_{jk}\left (1-\frac{a_{ki}}{a_{ji}}\right ), \\
\, \\
A_{ki}=a_{ki}+\partial_i\ln \left(1-\frac{a_{ki}}{a_{ji}}\right), \\
\, \\
A_{jk}=a_{jk}+\partial_k\ln A_{ij}, \\
\, \\
A_{kj}=a_{kj}+\partial_j\ln \left(1-\frac{a_{ki}}{a_{ji}}\right), \\
\, \\
A_{kl}=a_{kl}+\partial_l\ln \left(1-\frac{a_{ki}}{a_{ji}}\right),
\end{array} \eqno(10)
$$
where $k, l\ne i, j$ (compare with \cite{Kamran}, \cite{Kamran1}).
The nonsymmetry of these expressions is due to
the distinguished role played by the indices $i$ and $j$ in the
definition of the Laplace transformation $S_{ij}$.
Laplace transformations preserve the semihamiltonian property
and hence map integrable systems to integrable.

In the case $n=2$ there
is actually only one transformation
$S_{12}$ since $S_{12}\circ S_{21}=id$.

Some further properties of transformations $S_{ij}$ are discussed in sect. 3,
where we propose also a complete description of quadruples of $3\times 3$
systems, which are closed under all Laplace transformations $S_{ij}$.

In section 4  formulas for Laplace transformations of the
Lame and rotation coefficients are presented.

\section{Laplace transformations of two-component systems. Sequences of period
2.}

Iterating Laplace transformations according to formulas (5) we arrive at
the infinite sequence of systems with the characteristic velocities being
consequtive solutions of the integrable chain
$$
\frac{\partial_1\partial_2\lambda^n}{\partial_1\lambda^n\partial_2\lambda^n}
=\frac{1}{\lambda^n-\lambda^{n-1}} + \frac{1}{\lambda^n-\lambda^{n+1}},
\eqno(11)
$$
so that the Laplace transformation of the system with
characteristic velocities \linebreak
$(\lambda^{n-1}, \lambda^n)$ is the system with characteristic velocities
$(\lambda^{n}, \lambda^{n+1})$. In terms of the coefficients
$$
a^n=\frac{\partial_2\lambda^n}{\lambda^{n+1}-\lambda^n}, ~~~
b^n=\frac{\partial_1\lambda^n}{\lambda^{n-1}-\lambda^n},
$$
the chain (11) assumes the form
$$
\partial_1a^n=a^n(b^{n+1}-b^n), ~~~~ \partial_2b^n=b^n(a^{n-1}-a^n),
$$
and after the substitution 
$$
q^n=\ln a^nb^n =\ln {\frac{\partial_2\lambda^n \partial_1\lambda^n}
{(\lambda^{n+1}-\lambda^n)(\lambda^{n-1}-\lambda^n)}}
$$
reduces to  the well-known Toda chain
$$
\partial_1\partial_2q^n=2e^{q^n}-e^{q^{n+1}}-e^{q^{n-1}}. \eqno(12)
$$
Chain (11) appeared recently in \cite{ShabYam} 
as a symmetry of the
(2+1)-dimensional integrable equations of Davey-Stewartson type - see sect. 5.

A number of interesting results in the theory of Laplace transformations
have been derived while studying periodic sequences. It was demonstrated in
\cite{VesNov}, that periodic sequences
are ultimately connected with the spectral
theory of the two-dimensional Shroedinger operator. It turns out, that
any periodic sequence
of Laplace transformations of second order equations (2)
can be pulled back to the periodic sequence of Laplace transformations of
the corresponding hydrodynamic type systems. This pull-back is governed by
periodic reductions of chain (11).

{\bf Example.} Let us consider periodic sequence of Laplace transformations
of period 2, where equation (2) with coefficients $(a,b)$
transforms first into  equation
(2) with coefficients $(A, B)$, and than back into $(a, b)$.
It follows from (3) that $Aa=\varphi_2(R^2), ~~ Bb=\varphi_1(R^1)$,
where the functions $\varphi _i$ can be reduced to $\pm 1$ by a change to the
new Riemann invariants, so that we can assume
$Aa=\pm 1, ~~ Bb=\pm 1$. In what follows we consider the case
$A=\frac{1}{a}, ~~ B=\frac{1}{b}$ so that our Laplace sequence can be
schematically represented as follows:
$$
(a, b)\to (\frac{1}{a}, \frac{1}{b})\to (a, b). \eqno(13)
$$
Moreover, the coefficients $a$ and $b$ must obey the equations
$$
\partial_2\ln b =a-\frac{1}{a}, ~~~
\partial_1\ln a =b-\frac{1}{b}, \eqno(14)
$$
which are equivalent to the sh-Gordon equation
$$
\partial_1\partial_2\varphi =4 sh~\varphi,
$$
for $\varphi = \ln ab$. The periodic pull-back, corresponding to
the sequence (13), is of the form
$$
(\lambda^1, \lambda^2)\to
(\Lambda^1, \Lambda^2)\to
(\lambda^1, \lambda^2),
$$
where
$\Lambda^1=\lambda^2, ~~ \Lambda^2=\lambda^1$,
as follows automatically from (5).
Here  $a, b, \lambda^1, \lambda^2$ are connected by the formulas
$$
\begin{array}{c}
\frac{\partial_2\lambda^1}{\lambda^2-\lambda^1}=a, ~~~~
\frac{\partial_1\lambda^2}{\lambda^1-\lambda^2}=b,\\
\ \\
\frac{\partial_2\lambda^2}{\lambda^1-\lambda^2}=\frac{1}{a}, ~~~~
\frac{\partial_1\lambda^1}{\lambda^2-\lambda^1}=\frac{1}{b},
\end{array} \eqno(15)
$$
which are compatible due to (14).
One can show that $\lambda^1, \lambda^2$ satisfy
the second order system
$$
\frac{\partial_1\partial_2\lambda^1}{\partial_1\lambda^1\partial_2\lambda^1}
=\frac{2}{\lambda^1-\lambda^2}, ~~~~~~
\frac{\partial_1\partial_2\lambda^2}{\partial_1\lambda^2\partial_2\lambda^2}
=\frac{2}{\lambda^2-\lambda^1},
$$
which is just periodic reduction of  chain (11) of period 2:
~ $\lambda^3=\lambda^1, ~\lambda^4=\lambda^2$ and can be obtained by varying
the Lagrangian
$$
L=\int \int \frac{\partial_1\lambda^1\partial_2\lambda^2}
{(\lambda^1-\lambda^2)^2}~dR^1dR^2.
$$
Formulas (15) describe periodic pull-back of Laplace sequence
of period 2. This pull-back is defined uniquely up to transformations
$\lambda^1\to p\lambda^1+q, ~~
 \lambda^2\to p\lambda^2+q, ~~ p, q = const.$
Let us point out that applying Laplace transformations to commuting flows
with the characteristic velocities $w^1, w^2$, which do not satisfy the
restrictions (15) (although correspond to the same $a, b$),
we will not return back after going round the cycle.

{\bf Remark.} For any periodic sequence
of Laplace transformations of period two
$$
(a, b)\to (\frac{1}{a}, \frac{1}{b})\to (a, b)
$$
one can construct nonperiodic pull-back of the form
$$
(w^1, w^2)\to (w^2, \mu w^1)\to (\mu w^1, \mu w^2),
$$
where $\mu = const$. Here $a, b, w^1, w^2$ are connected by the formulas
$$
\begin{array}{c}
\frac{\partial_2w^1}{w^2-w^1}=a, ~~~~
\frac{\partial_1w^2}{w^1-w^2}=b,\\
\ \\
\frac{\partial_2w^2}{\mu w^1-w^2}=\frac{1}{a}, ~~~~
\frac{\mu \partial_1w^1}{w^2-\mu w^1}=\frac{1}{b}.
\end{array} \eqno(16)
$$
The linear system (16) is compatible and manifests the spectral problem
for equations (14) with the spectral parameter $\mu $.
It follows from (16) that $(\lambda^1, \lambda^2)$ are the components of
the wave function $(w^1, w^2)$ at the point $\mu =1$.

The situation with periodic sequences of an arbitrary period $n$
is completely analogous, namely,
for any periodic sequence of Laplace transformations of
second order equations of period
$n$ there exists exactly $n$-parameter family of periodic sequences of
systems of hydrodynamic type with the same period.
In a similar way (considering nonperiodic
pull-backs) one can construct spectral problems, corresponding to
periodic sequences of Laplace transformations of an arbitrary period $n$.

\bigskip
Let us discuss now another  question, concerning Laplace 
sequences of period two
$$
(a, b)\to (\frac{1}{a}, \frac{1}{b})\to (a, b),
$$
namely, the existence of  periodic sequences of solutions.
Applying two consequtive Laplace transformations in the direction $R^1$
to the initial solution $u$ of equation (2)
$$
\partial_1\partial_2u=a\partial_1u+b\partial_2u,
$$
and keeping in mind the conditions $A=\frac{1}{a}, ~ B=\frac{1}{b}$
(see the example), we obtain a new solution
$$
\widehat L_1 (u)=\partial_1^2u-(\frac{\partial_1b}{b}+b+\frac{1}{b})
\partial_1u+u,
$$ 
which does not necessarily coincide with $u$. In this sense transformation
$\widehat L_1$ is a recursion operator for the equation (2). 
In a similar way one can construct another recursion operator
$$
\widehat L_2(u)=\partial_2^2u-(\frac{\partial_2a}{a}+a+\frac{1}{a})
\partial_2u+u,
$$ 
generated by two consequtive Laplace transformations in the direction $R^2$.
It is an easy exercise to show that in general there are no solutions of
period 2, that is, solutions, satisfying any of the equivalent conditions
$\widehat L_1(u)=u$ or $\widehat L_2(u)=u$. So we will look for  solutions of
period 4, which obviously can be characterized by the constraint 
$\widehat L_1(u)=\widehat L_2(u)$ or, equivalently,
$$
\begin{array}{c}
\partial_1^2u=(\frac{\partial_1b}{b}+b+\frac{1}{b})\partial_1u+p,\\
\ \\
\partial_2^2u=(\frac{\partial_2a}{a}+a+\frac{1}{a})\partial_2u+p,\\
\end{array}
\eqno(2')
$$ 
for appropriate $p$.
Writing down the compatibility conditions of $(2')$ with $(2)$ 
and keeping in mind (14) we obtain the following equations for $p$:
$$
\begin{array}{c}
\partial_1p=\frac{b}{a}\partial_2u-\partial_1u+bp,\\
\ \\
\partial_2p=\frac{a}{b}\partial_1u-\partial_2u+ap.\\
\end{array}
\eqno(2'')
$$
Moreover, the compatibility conditions of $(2'')$ are satisfied identically.
Hence any periodic sequence of Laplace equations of period 2 possesses exactly
3-dimensional space of periodic solutions of period 4, which are described by
an involutive system $(2)$, $(2')$, $(2'')$. Three linearly independent solutions of
this system define a surface $M^2$ in 3-space, parametrized by coordinates 
$R^1, R^2$. This coordinate net is conjugate
due to (2) and generates  periodic Laplace sequence consisting of 4 surfaces
 $M^2\to M^2_1\to M^2_2 \to M^2_3 \to M^2$
in the standard differential-geometric sense -- see e.g.
\cite{Bol} and references therein for the properties of Laplace sequences of
period 4. 
One can also show, that the second quadratic forms of 
the surface $M^2$ and it's
Laplace images have the special isothermic form in the coordinates
$R^1, R^2$: they are proportional to $(dR^1)^2+(dR^2)^2$,  
so that all  congruences, generating this Laplace sequence, 
are W-congruences, and the corresponding conjugate nets are the so-called 
R-nets. 
Since the radius-vectors of $M^2$ and $M^2_2$ satisfy one and the same 
linear system $(2)$, $(2')$, $(2'')$,
these surfaces differ only by an affine
transformation of the 3-space (although do not coincide). The same property 
holds for  $M^2_1$ and $M^2_3$.
These considerations can be generalised to the case of arbitrary period $n$ as 
follows: any periodic equation (2) of  an arbitrary period $n$ possesses 
$(n-1)$-dimensional space of solutions with the same period $n$ -- see e.g.
\cite{Tzitzeica}.

Let us show, that any system of hydrodynamic type, which satisfies 
equations (15) and generates Laplace sequence of period two
$$
(\lambda^1, \lambda^2)\to
(\lambda^2, \lambda^1)\to
(\lambda^1, \lambda^2),
$$
possesses a unique conservative representation 
$$
\begin{array}{c}
u^1_t=f^1_x, \\
u^2_t=f^2_x
\end{array}
$$
of period 4. For that purpose we have to choose the densities $u$ and 
the fluxes $f$ in
such a way that $\widehat L_1(u)=\widehat L_2(u)$  and 
$\widehat L_1(f)=\widehat L_2(f)$, where
$$
\widehat L_1 (f)=\lambda^1\partial_1^2u-
(\frac{\lambda^1\partial_1b}{b}+\lambda^1b+\frac{\lambda^2}{b})
\partial_1u+f,
$$ 
$$
\widehat L_2 (f)=\lambda^2\partial_2^2u-
(\frac{\lambda^2\partial_2a}{b}+\lambda^2a+\frac{\lambda^1}{a})
\partial_2u+f,
$$ 
according to the transformation law of the fluxes, see Theorem 1.
Finally we obtain the following system for the conserved densities $u$
$$
\begin{array}{c}
\partial_1\partial_2u=a\partial_1u+b\partial_2u, \\
\ \\
\partial_1^2u=(\frac{\partial_1b}{b} + b)\partial_1u-
\frac{\partial_2u}{a}, \\
\ \\
\partial_2^2u=(\frac{\partial_2a}{a} + a)\partial_2u-
\frac{\partial_1u}{b}, \\
\end{array}
$$
which can be obtained from $(2)$, $(2')$, $(2'')$ by the reduction
$p=-\frac{\partial_1u}{b}-\frac{\partial_2u}{a}$. This system
is compatible and defines a 2-parameter space of conserved densities, 
providing together with the corresponding fluxes the unique conservative 
representation of period 4.

\section{Laplace transformations of $n$-component systems}

{\bf Lemma.} {\it Transformations $S_{ij}$ satisfy the identities}
$$
\begin{array}{c}
S_{ij}\circ S_{ji}=id, \\
S_{ij}=S_{ik}\circ S_{kj}=S_{kj}\circ S_{ik}, ~~~ k\ne i, j.
\end{array} \eqno(17)
$$
The proof can be obtained by a direct calculation.
The analogous identities are well-known in the theory of
$n$-conjugate coordinate systems -- see, e.g. \cite{Darboux}, p. 275.
>From (17) it immediately follows, that transformations $S_{ij}$
form a free abelian group with $n-1$ generators (one can
take e.g. $S_{12},...,S_{1n}$ as the generators).

In the language of
$n$-conjugate coordinate systems transformations $S_{ij}$
have been discussed in papers \cite{Chern},
\cite{Koz'mina}, \cite{Smirnov}, \cite{Shul'man}, \cite{AkGold},
which partially duplicate the investigations of
Darboux \cite{Darboux}, p. 274-275. The summary of these results can be found in the book
\cite{AkGold1}.
Higher dimensional Laplace invariants and terminating Laplace
sequences have been investigated recently in \cite{Kamran}, \cite{Kamran1}, see also
\cite{Eisenhar}, \cite{Korovin}.
Let us also point to the paper
\cite{Athorne}, where the method of factorization has been successfully
applied to construct Laplace transformations and Laplace invariants of
multidimensional matrix differential operators of the first order.

It looks promising to continue the investigation of transformations
$S_{ij}$, in particular:

-- investigate finite families of hydrodynamic type systems, which are
closed under all Laplace transformations $S_{ij}$
(analogs of closed Laplace sequences in the case $n=2$).
It looks likely, that these systems should enjoy the property of certain
"extra" integrability;

-- study the behaviour of Hamiltonian structures under the Laplace
transformations $S_{ij}$. We emphasize that local Hamiltonian structures of
Dubrovin-Novikov type \cite{DubNov} are not preserved under the Laplace transformations.

Let us give now a description of quadruples of
$3\times 3$ hydrodynamic type systems, which are closed under
all Laplace transformations.
Let the characteristic velocities of the systems
$\Sigma$, ~ $\Sigma_1$, ~ $\Sigma_2$, ~ $\Sigma_3$ be respectively
$(\lambda^1, \lambda^2, \lambda^3)$, ~
$(\lambda^4, \lambda^3, \lambda^2)$, ~
$(\lambda^3, \lambda^4, \lambda^1)$, ~
$(\lambda^2, \lambda^1, \lambda^4)$
 -- see the picture.

\vspace{2truecm}
\hspace{-1.5truecm}
\unitlength 1.00 mm
\thicklines
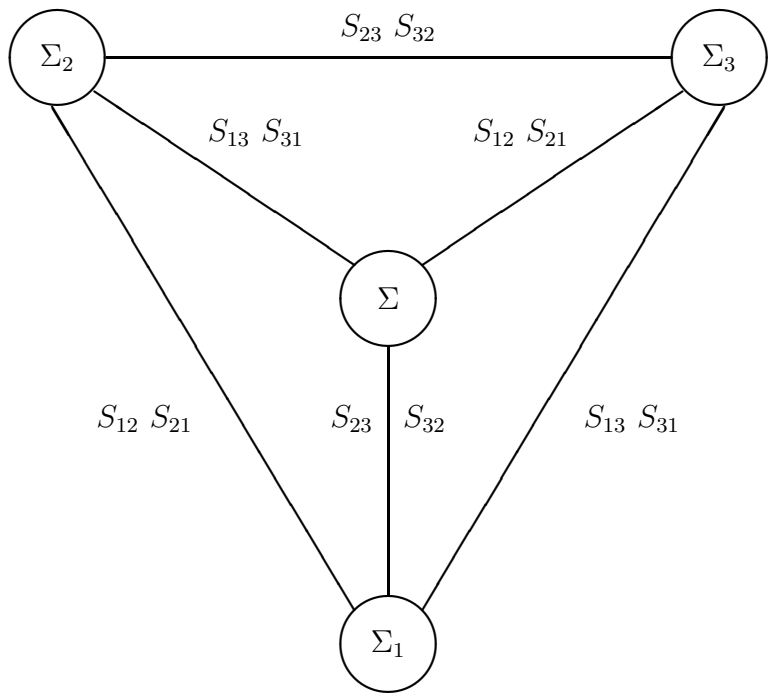
\begin{picture}(130.00,110.00)
\put(36.00,104.00){\circle{12.00}}
\put(124.00,104.00){\circle{12.00}}
\put(80.00,72.00){\circle{12.00}}
\put(80.00,26.00){\circle{12.00}}
\put(42.50,104.00){\line(1,0){75.00}}
\put(80.00,32.40){\line(0,1){33.30}}
\put(84.50,76.50){\line(3,2){34.60}}
\put(75.50,76.50){\line(-3,2){34.60}}
\put(84.50,30.50){\line(3,5){40.20}}
\put(75.50,30.50){\line(-3,5){40.20}}
\put(80.00,72.00){\makebox(0,0)[cc]{$\Sigma$}}
\put(124.00,104.00){\makebox(0,0)[cc]{$\Sigma_3$}}
\put(36.00,104.00){\makebox(0,0)[cc]{$\Sigma_2$}}
\put(80.00,26.00){\makebox(0,0)[cc]{$\Sigma_1$}}
\put(106.00,56.00){\makebox(0,0)[lc]{$S_{13}$  $S_{31}$}}
\put(54.00,56.00){\makebox(0,0)[rc]{$S_{12}$ $S_{21}$}}
\put(78.00,56.00){\makebox(0,0)[rc]{$S_{23}$}}
\put(82.00,56.00){\makebox(0,0)[lc]{$S_{32}$}}
\put(56.00,94.00){\makebox(0,0)[lc]{$S_{13}$ $S_{31}$}}
\put(104.00,94.00){\makebox(0,0)[rc]{$S_{12}$ $S_{21}$}}
\put(80.00,108.00){\makebox(0,0)[cc]{$S_{23}$ $S_{32}$}}
\end{picture}

The marked lines joining pairs of systems on the picture indicate,
for instance, that  system $\Sigma_3$ can be obtained from  $\Sigma$
by the Laplace transformations $S_{12}$ and $S_{21}$
(vice-versa, system $\Sigma$ can be obtained
from $\Sigma_3$ by transformations $S_{12}$ and $S_{21}$).
The lines are marked in accordance with the identities (17).
As far as in our construction $S_{ij}^2=id$ for any pair of indices $i, j$,
the quadruples of systems under consideration are complete analogs
of closed Laplace sequences of period 2.
Formulas (8) result in the complicated overdetermined system
for the characteristic velocities
$\lambda^1$ -- $\lambda^4$:
$$
\begin{array}{c}
\frac{\partial_1\partial_2\lambda^1}{\partial_1\lambda^1\partial_2\lambda^1}=
\frac{2}{\lambda^1-\lambda^2}, ~~~
\frac{\partial_1\partial_2\lambda^2}{\partial_1\lambda^2\partial_2\lambda^2}=
\frac{2}{\lambda^2-\lambda^1}, \\
\ \\
\frac{\partial_1\partial_3\lambda^1}{\partial_1\lambda^1\partial_3\lambda^1}=
\frac{2}{\lambda^1-\lambda^3}, ~~~
\frac{\partial_1\partial_3\lambda^3}{\partial_1\lambda^3\partial_3\lambda^3}=
\frac{2}{\lambda^3-\lambda^1}, \\
\ \\
\frac{\partial_2\partial_3\lambda^2}{\partial_2\lambda^2\partial_3\lambda^2}=
\frac{2}{\lambda^2-\lambda^3}, ~~~
\frac{\partial_2\partial_3\lambda^3}{\partial_2\lambda^3\partial_3\lambda^3}=
\frac{2}{\lambda^3-\lambda^2},  \\
\ \\
\frac{\partial_1\partial_2\lambda^3}{\partial_1\lambda^3\partial_2\lambda^3}=
\frac{2}{\lambda^3-\lambda^4}, ~~~
\frac{\partial_1\partial_2\lambda^4}{\partial_1\lambda^4\partial_2\lambda^4}=
\frac{2}{\lambda^4-\lambda^3}, \\
\ \\
\frac{\partial_1\partial_3\lambda^2}{\partial_1\lambda^2\partial_3\lambda^2}=
\frac{2}{\lambda^2-\lambda^4}, ~~~
\frac{\partial_1\partial_3\lambda^4}{\partial_1\lambda^4\partial_3\lambda^4}=
\frac{2}{\lambda^4-\lambda^2}, \\
\ \\
\frac{\partial_2\partial_3\lambda^1}{\partial_2\lambda^1\partial_3\lambda^1}=
\frac{2}{\lambda^1-\lambda^4}, ~~~
\frac{\partial_2\partial_3\lambda^4}{\partial_2\lambda^4\partial_3\lambda^4}=
\frac{2}{\lambda^4-\lambda^1},
\end{array} \eqno(18)
$$
and
$$
\begin{array}{c}
\partial_1\lambda^2\frac{\lambda^4-\lambda^3}{\lambda^1-\lambda^2}=
\partial_1\lambda^3\frac{\lambda^4-\lambda^2}{\lambda^1-\lambda^3}, ~~~~
\partial_1\lambda^1\frac{\lambda^2-\lambda^4}{\lambda^3-\lambda^1}=
\partial_1\lambda^4\frac{\lambda^2-\lambda^1}{\lambda^3-\lambda^4},  \\
\ \\
\partial_2\lambda^1\frac{\lambda^4-\lambda^3}{\lambda^2-\lambda^1}=
\partial_2\lambda^3\frac{\lambda^4-\lambda^1}{\lambda^2-\lambda^3}, ~~~~
\partial_2\lambda^2\frac{\lambda^1-\lambda^4}{\lambda^3-\lambda^2}=
\partial_2\lambda^4\frac{\lambda^1-\lambda^2}{\lambda^3-\lambda^4},  \\
\ \\
\partial_3\lambda^1\frac{\lambda^4-\lambda^2}{\lambda^3-\lambda^1}=
\partial_3\lambda^2\frac{\lambda^4-\lambda^1}{\lambda^3-\lambda^2}, ~~~~
\partial_3\lambda^3\frac{\lambda^1-\lambda^4}{\lambda^2-\lambda^3}=
\partial_3\lambda^4\frac{\lambda^1-\lambda^3}{\lambda^2-\lambda^4}.
\end{array} \eqno(19)
$$
It follows from (19) that the cross-ratio of 4 characteristic velocities
$\lambda^1$ -- $\lambda^4$ is constant:
$$
\frac{(\lambda^1-\lambda^2)(\lambda^3-\lambda^4)}
{(\lambda^1-\lambda^4)(\lambda^3-\lambda^2)}=\mu=const.
$$
Excluding $\lambda^4$ one can rewrite equations (18),
(19) in a simplified form
$$
\begin{array}{c}
\frac{\partial_1\partial_2\lambda^1}{\partial_1\lambda^1\partial_2\lambda^1}=
\frac{2}{\lambda^1-\lambda^2}, ~~~
\frac{\partial_1\partial_2\lambda^2}{\partial_1\lambda^2\partial_2\lambda^2}=
\frac{2}{\lambda^2-\lambda^1}, \\
\ \\
\frac{\partial_1\partial_3\lambda^1}{\partial_1\lambda^1\partial_3\lambda^1}=
\frac{2}{\lambda^1-\lambda^3}, ~~~
\frac{\partial_1\partial_3\lambda^3}{\partial_1\lambda^3\partial_3\lambda^3}=
\frac{2}{\lambda^3-\lambda^1}, \\
\ \\
\frac{\partial_2\partial_3\lambda^2}{\partial_2\lambda^2\partial_3\lambda^2}=
\frac{2}{\lambda^2-\lambda^3}, ~~~
\frac{\partial_2\partial_3\lambda^3}{\partial_2\lambda^3\partial_3\lambda^3}=
\frac{2}{\lambda^3-\lambda^2},
\end{array}  \eqno(20)
$$
and
$$
\begin{array}{c}
\mu(\lambda^1-\lambda^3)^2\partial_1\lambda^2=
(\mu-1)(\lambda^1-\lambda^2)^2\partial_1\lambda^3, \\
\ \\
\mu(\lambda^2-\lambda^3)^2\partial_2\lambda^1=
(\lambda^1-\lambda^2)^2\partial_2\lambda^3, \\
\ \\
(1-\mu)(\lambda^2-\lambda^3)^2\partial_3\lambda^1=
(\lambda^1-\lambda^3)^2\partial_3\lambda^2,
\end{array} \eqno(21)
$$
so that (18), (19) are equivalent to
(20), (21). It follows from (20), that
$$
\begin{array}{c}
\partial_1\lambda^1\partial_1\lambda^2=(\lambda^1-\lambda^2)^2
\varphi_{13}, ~~~~
\partial_2\lambda^1\partial_2\lambda^2=(\lambda^1-\lambda^2)^2
\varphi_{23}, \\
\ \\
\partial_1\lambda^1\partial_1\lambda^3=(\lambda^1-\lambda^3)^2\eta_{12}, ~~~~
\partial_3\lambda^1\partial_3\lambda^3=(\lambda^1-\lambda^3)^2\eta_{23}, \\
\ \\
\partial_2\lambda^2\partial_2\lambda^3=(\lambda^2-\lambda^3)^2s_{12}, ~~~~
\partial_3\lambda^2\partial_3\lambda^3=(\lambda^2-\lambda^3)^2s_{13},
\end{array} \eqno(22)
$$
where $\varphi_{ij}(R^i, R^j), ~ \eta_{ij}(R^i, R^j), ~ s_{ij}(R^i, R^j)$
are arbitrary functions of the specified arguments.
Using (21) one immediately  arrives at the following relations between
$\varphi_{ij}, ~ \eta_{ij}, ~ s_{ij}$:
$$
\mu\varphi_{13}=(\mu-1)\eta_{12}, ~~~
\mu\varphi_{23}=s_{12}, ~~~ (1-\mu)\eta_{23}=s_{13},
$$
so that
$$
\begin{array}{c}
\varphi_{13}=(\mu-1)\varphi_1(R^1), ~~~ \varphi_{23}=\varphi_2(R^2), \\
\ \\
\eta_{12}=\mu\varphi_1(R^1), ~~~ \eta_{23}=\varphi_3(R^3), \\
\ \\
s_{12}=\mu\varphi_2(R^2), ~~~ s_{13}=(1-\mu)\varphi_3(R^3),
\end{array}
$$
where $\varphi_i(R^i)$  are arbitrary functions,
which can be reduced to $\pm 1$ by the appropriate change of Riemann
invariants. In what follows we consider the case $\varphi_i=1$, so that
equations (22) assume the form
$$
\begin{array}{c}
\partial_1\lambda^1\partial_1\lambda^2=(\mu-1)(\lambda^1-\lambda^2)^2, ~~~~
\partial_2\lambda^1\partial_2\lambda^2=(\lambda^1-\lambda^2)^2, \\
\ \\
\partial_1\lambda^1\partial_1\lambda^3=\mu(\lambda^1-\lambda^3)^2, ~~~~
\partial_3\lambda^1\partial_3\lambda^3=(\lambda^1-\lambda^3)^2, \\
\ \\
\partial_2\lambda^2\partial_2\lambda^3=\mu(\lambda^2-\lambda^3)^2, ~~~~
\partial_3\lambda^2\partial_3\lambda^3=(1-\mu)(\lambda^2-\lambda^3)^2.
\end{array} \eqno(23)
$$
We emphasize that (20), (21) are equivalent to (23).
Let us demonstrate, that equations (23) are equivalent to a special quadratic
reduction of the $(2+1)$-dimensional $3$-wave system. For that purpose we
introduce differential  1-forms
$$
\begin{array}{c}
\omega^1=\sqrt{\frac{\mu}{\mu-1}}
\frac{\lambda^1-\lambda^3}{(\lambda^1-\lambda^2)(\lambda^2-\lambda^3)}~
d\lambda^2 -
\sqrt{\frac{\mu-1}{\mu}}
\frac{\lambda^1-\lambda^2}{(\lambda^1-\lambda^3)(\lambda^2-\lambda^3)}~
d\lambda^3, \\
\ \\
\omega^2=\sqrt{\mu}
\frac{\lambda^2-\lambda^3}{(\lambda^1-\lambda^2)(\lambda^1-\lambda^3)}~
d\lambda^1 -
\frac{1}{\sqrt{\mu}}
\frac{\lambda^1-\lambda^2}{(\lambda^1-\lambda^3)(\lambda^2-\lambda^3)}~
d\lambda^3, \\
\ \\
\omega^3=-\sqrt{\mu-1}
\frac{\lambda^2-\lambda^3}{(\lambda^1-\lambda^2)(\lambda^1-\lambda^3)}~
d\lambda^1 -
\frac{1}{\sqrt{\mu-1}}
\frac{\lambda^1-\lambda^3}{(\lambda^1-\lambda^2)(\lambda^2-\lambda^3)}~
d\lambda^2,
\end{array} \eqno(24)
$$
which are choosen in such a way that equations (21)
become just
$$
\begin{array}{c}
\omega^1\wedge dR^2\wedge dR^3=0, \\
\omega^2\wedge dR^1\wedge dR^3=0, \\
\omega^3\wedge dR^1\wedge dR^2=0.
\end{array} \eqno(25)
$$
Moreover, the forms $\omega^i$  satisfy the structure equations of the
Lie group $SO(2, 1)$:
$$
\begin{array}{c}
d\omega^1=\omega^2\wedge \omega^3, ~~~
d\omega^2=\omega^3\wedge \omega^1, ~~~
d\omega^3=\omega^2\wedge \omega^1.
\end{array} \eqno(26)
$$
Let us introduce the coefficients  $\beta_{ij}$ by the formulas
$$
\begin{array}{c}
\omega^1=\beta_{32}dR^2-\beta_{23}dR^3, \\
\omega^2=\beta_{13}dR^3-\beta_{31}dR^1, \\
\omega^3=\beta_{12}dR^2-\beta_{21}dR^1,
\end{array} \eqno(27)
$$
(the validity of such representation is due to (25)).
Coefficients $\beta_{ij}$ satisfy the nonlinear system,
which is well-known in the theory of $3$-orthogonal
coordinates. This system results from the substitution of (27)
into the structure equations (26):
$$
\begin{array}{c}
\partial_1\beta_{23}=-\beta_{21}\beta_{13}, ~~~
\partial_1\beta_{32}=-\beta_{31}\beta_{12}, \\
\partial_2\beta_{13}=-\beta_{12}\beta_{23}, ~~~
\partial_2\beta_{31}=-\beta_{32}\beta_{21}, \\
\partial_3\beta_{12}=\beta_{13}\beta_{32}, ~~~
\partial_3\beta_{21}=\beta_{23}\beta_{31},
\end{array} \eqno(28)
$$
and
$$
\begin{array}{c}
\partial_1\beta_{12}+\partial_2\beta_{21}+\beta_{31}\beta_{32}=0, \\
\partial_1\beta_{13}+\partial_3\beta_{31}-\beta_{21}\beta_{23}=0, \\
\partial_2\beta_{23}+\partial_3\beta_{32}-\beta_{12}\beta_{13}=0.
\end{array} \eqno(29)
$$
The transformation from $\lambda^i$  to $\beta_{ij}$
is just the differential substitution of the first order.
The explicit expressions for  $\beta_{ij}$  can be obtained by comparing
(27) and (24):
$$
\begin{array}{c}
\beta_{32}=\sqrt{\frac{\mu}{\mu-1}}
\frac{\lambda^1-\lambda^3}{(\lambda^1-\lambda^2)(\lambda^2-\lambda^3)}~
\partial_2\lambda^2 -
\sqrt{\frac{\mu-1}{\mu}}
\frac{\lambda^1-\lambda^2}{(\lambda^1-\lambda^3)(\lambda^2-\lambda^3)}~
\partial_2\lambda^3, \\
\ \\
\beta_{23}=-\sqrt{\frac{\mu}{\mu-1}}
\frac{\lambda^1-\lambda^3}{(\lambda^1-\lambda^2)(\lambda^2-\lambda^3)}~
\partial_3\lambda^2 +
\sqrt{\frac{\mu-1}{\mu}}
\frac{\lambda^1-\lambda^2}{(\lambda^1-\lambda^3)(\lambda^2-\lambda^3)}~
\partial_3\lambda^3, \\
\ \\
\beta_{13}=\sqrt{\mu}
\frac{\lambda^2-\lambda^3}{(\lambda^1-\lambda^2)(\lambda^1-\lambda^3)}~
\partial_3\lambda^1 -
\frac{1}{\sqrt{\mu}}
\frac{\lambda^1-\lambda^2}{(\lambda^1-\lambda^3)(\lambda^2-\lambda^3)}~
\partial_3\lambda^3, \\
\ \\
\beta_{31}=-\sqrt{\mu}
\frac{\lambda^2-\lambda^3}{(\lambda^1-\lambda^2)(\lambda^1-\lambda^3)}~
\partial_1\lambda^1 +
\frac{1}{\sqrt{\mu}}
\frac{\lambda^1-\lambda^2}{(\lambda^1-\lambda^3)(\lambda^2-\lambda^3)}~
\partial_1\lambda^3, \\
\ \\
\beta_{12}=-\sqrt{\mu-1}
\frac{\lambda^2-\lambda^3}{(\lambda^1-\lambda^2)(\lambda^1-\lambda^3)}~
\partial_2\lambda^1 -
\frac{1}{\sqrt{\mu-1}}
\frac{\lambda^1-\lambda^3}{(\lambda^1-\lambda^2)(\lambda^2-\lambda^3)}~
\partial_2\lambda^2, \\
\ \\
\beta_{21}=\sqrt{\mu-1}
\frac{\lambda^2-\lambda^3}{(\lambda^1-\lambda^2)(\lambda^1-\lambda^3)}~
\partial_1\lambda^1 +
\frac{1}{\sqrt{\mu-1}}
\frac{\lambda^1-\lambda^3}{(\lambda^1-\lambda^2)(\lambda^2-\lambda^3)}~
\partial_1\lambda^2.
\end{array}
$$
Moreover, equations (23) impose
the following quadratic reduction on the coefficients
$\beta_{ij}$:
$$
\frac{\beta_{32}^2}{\mu}-\beta_{12}^2=-4, ~~~~
\frac{\beta_{23}^2}{\mu-1}-\beta_{13}^2=4, ~~~~
\frac{\beta_{21}^2}{\mu-1}-\frac{\beta_{31}^2}{\mu}=4.
\eqno(30)
$$
Let us point out, that equations (28) are just $(2+1)$-dimensional
$3$-wave system, corresponding to the spectral problem
$$
\begin{array}{c}
\partial_2H_1=-\beta_{21}H_2, ~~~~ \partial_3H_1=-\beta_{31}H_3, \\
\partial_1H_2=-\beta_{12}H_1, ~~~~ \partial_3H_2=-\beta_{32}H_3, \\
\partial_1H_3= \beta_{13}H_1, ~~~~ \partial_2H_3= \beta_{23}H_2.
\end{array}
$$
In order to comply with reduction (30) we introduce the parametrization
$$
\begin{array}{c}
\beta_{32}=2\sqrt{\mu}~sh~u, ~~~~ \beta_{12}=2~ch~u, \\
\beta_{23}=2\sqrt{\mu-1}~ch~v, ~~~~ \beta_{13}=2~sh~v, \\
\beta_{21}=2\sqrt{\mu-1}~ch~w, ~~~~ \beta_{31}=2\sqrt{\mu}~sh~w,
\end{array}
$$
so that equations (28) become
$$
\begin{array}{c}
\partial_1u=-2~sh~w, ~~~~ \partial_3u=2\sqrt{\mu}~sh~v, \\
\partial_1v=-2~ch~w, ~~~~ \partial_2v=-2\sqrt{\mu-1}~ch~u, \\
\partial_2w=-2\sqrt{\mu-1}~sh~u, ~~~~ \partial_3w=2\sqrt{\mu}~ch~v,
\end{array} \eqno(31)
$$
(we point out, that after this substitution equations (29)
are satisfied identically). After the appropriate rescaling of Riemann
invariants equations (31) assume the simple form
$$
\begin{array}{c}
\partial_1u=sh~w, ~~~~ \partial_3u=sh~v, \\
\partial_1v=ch~w, ~~~~ \partial_2v=ch~u, \\
\partial_2w=sh~u, ~~~~ \partial_3w=ch~v.
\end{array}
$$
Expressing $v$  and $w$ as follows
$$
v=arcsh~\partial_3u, ~~~~ w=arcsh~\partial_1u,
$$
one can rewrite this system in the form of
three pairwise commuting Monge-Ampere equations
$$
\begin{array}{c}
\partial_1\partial_2u=sh~u\sqrt{1+(\partial_1u)^2}, \\
\ \\
\partial_3\partial_2u=ch~u\sqrt{1+(\partial_3u)^2}, \\
\ \\
\partial_1\partial_3u=\sqrt{1+(\partial_1u)^2}\sqrt{1+(\partial_3u)^2}.
\end{array}
$$
(up to the authors knowledge the problem of classification of
commuting Monge-Ampere equations has not been addressed before).
It looks promising to continue the investigation of finite families
of $n\times n$ hydrodynamic type systems in Riemann invariants, which are
closed under all Laplace transformations $S_{ij}$.
It is naturall to restrict oneself to the case when $S_{ij}^k=id$
for some $k\geq 2$. Examples discussed above support the evidence that
the problem is nontrivial even in the simplest cases
$(n=2, k=2)$ and $(n=3, k=2)$.
The equations for the characteristic velocities of the
corresponding hydrodynamic type systems should reduce to appropriate
integrable reductions of the $(2+1)$-dimensional $n$-wave system
$$
\partial_k\beta_{ij}=\beta_{ik}\beta_{kj}.
$$

\section{Laplace transformations of the Lame and rotation coefficients}

In the 2-component case the Lame coefficients $h_1, h_2$ are defined by
the formulas
$$
\partial_2 \ln h_1=a, ~~~~ \partial_1\ln h_2=b.
\eqno(32)
$$
It turns out that Laplace transformation (3) can be pulled back to the
transformation of the Lame coefficients: the transformed Lame coefficients
$H_1, H_2$ are given by the formulas
$$
\begin{array}{c}
H_1=\frac{h_1}{b}=\frac{h_1h_2}{\partial_1h_2}, \\
\, \\
H_2=h_2A=h_2\partial_2 \ln H_1,
\end{array}
\eqno(33)
$$
(see Darboux \cite{Darboux}), so that
$$
\partial_2 \ln H_1=A, ~~~~ \partial_1 \ln H_2=B.
$$
In terms of the chain
$$
\begin{array}{c}
\partial_1\ln h^n_2 =\frac{h^n_1}{h^{n+1}_1}, \\
\, \\
\partial_2\ln h^n_1 =\frac{h^n_2}{h^{n-1}_2},
\end{array}
\eqno(34)
$$
transformation (33) reduces to the shift
$$
(h^n_1, h^n_2) \to (h^{n+1}_1, h^{n+1}_2).
$$
Formulas (33) can also be rewritten in terms of the rotation
coefficients
$$
\beta_{12}=\frac{\partial_1h_2}{h_1}, ~~~~
\beta_{21}=\frac{\partial_2h_1}{h_2},
\eqno(35)
$$
namely, the transformed rotation coefficients $\tilde \beta_{12}, 
\tilde \beta_{21}$ are the following:
$$
\begin{array}{c}
\tilde \beta_{12}=\frac{\partial_1H_2}{H_1}=
\beta_{12}(\beta_{12}\beta_{21}-\partial_1\partial_2\ln \beta_{12}),\\
\, \\
\tilde \beta_{21}=\frac{\partial_2H_1}{H_2}=\frac{1}{\beta_{12}}.
\end{array}
\eqno(36)
$$

The generalisation to the $n$-component case is straightforward: under
the transformation $S_{ij}$ the Lame coefficients $h_i$
defined by the formula
$$
\partial_j\ln h_i=a_{ij}
$$
transform into $H_i$ as follows:
$$
\begin{array}{c}
H_i=\frac{h_i}{a_{ji}}=\frac{h_ih_j}{\partial_ih_j},\\
\, \\
H_j=h_jA_{ij}=h_j\partial_j\ln H_i, \\
\, \\
H_k=h_k\left(1-\frac{a_{ki}}{a_{ji}}\right)=
h_k-\frac{\partial_ih_k}{\partial_ih_j}h_j, ~~~k\ne i, j.
\end{array}
\eqno(37)
$$
Indeed one can verify directly that
$$
\partial_j\ln H_i=A_{ij},
$$
where $A_{ij}$ are specified by (10).

In a similar way the rotation coefficients
$$
\beta_{ij}=\frac{\partial_ih_j}{h_i}
$$
transform into $\tilde \beta_{ij}$ as follows:
$$
\begin{array}{c}
\tilde \beta_{ij}=\beta_{ij}(\beta_{ij}\beta_{ji}-\partial_i\partial_j \ln
\beta_{ij}), \\
\, \\
\tilde \beta_{ji}=\frac{1}{\beta_{ij}}, \\
\, \\
\tilde \beta_{ik}=-\beta_{ij}\partial_i\frac{\beta_{ik}}{\beta_{ij}}, \\
\, \\
\tilde \beta_{ki}=\frac{\beta_{kj}}{\beta_{ij}}, \\
\, \\
\tilde \beta_{jk}=-\frac{\beta_{ik}}{\beta_{ij}}, \\
\, \\
\tilde \beta_{kj}=\beta_{ij}\partial_j\frac{\beta_{kj}}{\beta_{ij}}, \\
\, \\
\tilde \beta_{kl}=\beta_{kl}-\frac{\beta_{kj}\beta_{il}}{\beta_{ij}},
\end{array}
\eqno(38)
$$
where $ k, l \ne i, j$ (compare with \cite{Guichard}, p.12). In order to check 
that indeed
$$
\tilde \beta_{ij}=\frac{\partial_iH_j}{H_i},
$$
it is convenient to use the following equivalent representation for
the transformed rotation coefficients $H_i$:
$$
\begin{array}{c}
H_i=\frac{h_j}{\beta_{ij}},\\
\, \\
H_j=\partial_jh_j-h_j\partial_j\ln \beta_{ij}, \\
\, \\
H_k=h_k-\frac{\beta_{ik}}{\beta_{ij}}h_j, ~~~ k\ne i, j,
\end{array}
$$
and to keep in mind the equations
$$
\partial_k\beta_{ij}=\beta_{ik}\beta_{kj}, ~~~ i\ne j\ne k,
\eqno(39)
$$
satisfied by the rotation coefficients of an arbitrary
semihamiltonian system. It should be emphasized that the transformed
rotation coefficients satisfy the same equations (39)
and hence transformations
$S_{ij}$ written in the form (38) are just discrete symmetries of the
$(2+1)$-dimensional $n$-wave system (39).
The role of these transformations in the theory of $n$-wave system
deserves a special investigation. Basically transformations $S_{ij}$
preserve neither the Egorov reduction
$$
\beta_{ij}=\beta_{ji},
$$
nor the zero curvature reduction
$$
\partial_i\beta_{ij}+\partial_j\beta_{ji}+\sum_{k\ne i, j}
\beta_{ki}\beta_{kj}=0
$$
of the $n$-wave system (39).

{\bf Remark.} A way to generalize transformations $S_{ij}$ written
in the form (37), (38) is to allow the rescaling
$$
H_s \to \mu_s H_s, ~~~~~ \tilde \beta_{sl}\to \frac{\mu_l}{\mu_s}
\tilde \beta_{sl}, ~~~ \mu_s=const,
$$
where generically the scaling factors $\mu_s$ depend on $S_{ij}$.
One can always choose $\mu_s$ in such a way as to preserve the basic
identities (17). The main purpose for introducing the scaling factors
is the construction of solutions of the $n$-wave system (39) for which, say,
$S_{ij}^2=id$ (analogs of periodic sequences of period 2). For instance,
one can show that when $n\geq 3$ there are no nontrivial solutions of
(39) satisfying $S_{ij}^2=id$ if $S_{ij}$ are as in (38). However they do
exist if we rescale $S_{ij}$ appropriately.

\section{Laplace transformations as symmetries of the Davey-Stewartson-type
equations}

It should be noted, that Laplace transformation (5) for the characteristic 
velocities
$$
\begin{array}{c}
\Lambda^1=\lambda^2,  \\
\ \\
\Lambda^2=
\lambda^2-\frac{1}
{\frac{\partial_1\partial_2\lambda^2}
{\partial_1\lambda^2 \partial_2\lambda^2 }+
\frac{1}{\lambda^1-\lambda^2}}
\end{array}
$$
arises naturally as the symmetry of the (2+1)-dimensional integrable systems
$$
\begin{array}{c}
r_t=r_{xx}-\frac{2r_x^2}{r-s}+2r_xR, \\
\ \\
s_t=-s_{xx}-\frac{2s_x^2}{r-s}+2s_xR, \\
\ \\
R_y=\frac{r_xs_y-r_ys_x}{(r-s)^2},
\end{array}
$$
and
$$
\begin{array}{c}
r_t=r_{yy}-\frac{2r_y^2}{r-s}+2r_yS, \\
\ \\
s_t=-s_{yy}-\frac{2s_y^2}{r-s}+2s_yS, \\
\ \\
S_x=\frac{r_ys_x-r_xs_y}{(r-s)^2},
\end{array}
$$
which have been discussed recently in \cite{ShabYam} and are both equivalent
to the Davey-Stewartson equations.
These equations are invariant under the transformation
$$
\tilde r=s, ~~~
\tilde s=
s-\frac{1}
{\frac{s_{xy}}
{s_xs_y}+
\frac{1}{r-s}}, 
$$
if we define
$$
\tilde R=R+\left(\ln \frac{s_y}{s_x}\right)_x, ~~~~
\tilde S=S+\left(\ln \frac{s_x}{s_y}\right)_y.
$$
Formulas for $\tilde r, \tilde s$ 
coincide with (5) after introducing the notation
$$
r=\lambda^1, ~~ s=\lambda^2, ~~ \partial_1=\partial_x, ~~
\partial_2=\partial_y.
$$

In a similar way Laplace transformation (36) for the rotation coefficients
$$
\begin{array}{c}
\tilde \beta_{12}=
\beta_{12}(\beta_{12}\beta_{21}-\partial_1\partial_2\ln \beta_{12}),\\
\, \\
\tilde \beta_{21}=\frac{1}{\beta_{12}},
\end{array}
$$
arises as the symmetry of the Davey-Stewartson equations
$$
\begin{array}{c}
iu_t=u_{xx}+u_{yy}-u(p+q), \\
\ \\
-iv_t=v_{xx}+v_{yy}-v(p+q), \\
\ \\
p_x=2(uv)_y, ~~~ p_y=2(uv)_x,
\end{array}
$$
which were shown in \cite{Leznov} to be invariant under the transformation
$$
\begin{array}{c}
\tilde u=u(uv-{(\ln u)}_{xy}), \\
\tilde v=\frac{1}{u}, \\
\tilde p=p-2{(\ln u)}_{yy}, \\
\tilde q=q-2{(\ln u)}_{xx}.
\end{array}
$$
Formulas for $\tilde u, \tilde v$ coincide with (36) after introducing the 
notation
$$
\beta_{12}=u, ~~ \beta_{21}=v, ~~ \partial_1=\partial_x, ~~
\partial_2=\partial_y.
$$

\section{Geometric background.}

In this section we give geometric interpretation of formulas (5) based on
the correspondence between systems of conservation laws and line congruences
in the projective space.

Let us consider a $2\times 2$ system (1) in Riemann invariants
$$
\begin{array}{c}
R_t^1=\lambda^1(R)R^1_x,\\
R_t^2=\lambda^2(R)R^2_x,\\
\end{array}
$$
and choose any it's conservative representation
$$
\begin{array}{c}
u_t^1=f^1(u)_x,\\
u_t^2=f^2(u)_x,\\
\end{array}
\eqno(40)
$$
where $u=(u^1(R), u^2(R))$ are conserved densities of system (1) with the
corresponding fluxes $f=(f^1(R), f^2(R))$. We recall that the densities and the
fluxes satisfy the equations
$$
\partial_if=\lambda^i\partial_iu, ~~~ i=1, 2.
$$
Following \cite{AgaFer}, we associate with (40) a congruence of straight lines
in the 3-space $E^3(y^0, y^1, y^3)$ defined by the formulas
$$
\begin{array}{c}
y^1=u^1y^0-f^1, \\
y^2=u^2y^0-f^2.
\end{array}
\eqno(41)
$$
This correspondence was investigated in \cite{AgaFer}, where it was shown
that all familiar constructions in the theory of systems of conservation laws
(40) have their natural 
geometric counterpart in projective theory of congruences.
Let us recall the definition of the Laplace transformation of congruence (41).
Any congruence (41) has 2 focal surfaces with the radius-vectors
$\stackrel{\rightarrow }{r}$ and $\stackrel{\rightarrow }{R}$:
$$
\stackrel{\rightarrow }{r}=\left(
\begin{array}{c}
\lambda ^1 \\
\lambda ^1u^1-f^1 \\
\lambda ^1u^2-f^2
\end{array}
\right), ~~~~
\stackrel{\rightarrow }{R}=\left(
\begin{array}{c}
\lambda ^2 \\
\lambda ^2u^1-f^1 \\
\lambda ^2u^2-f^2
\end{array}
\right).
$$
The curves  $R^2=const$ and $R^1=const$ are conjugate on both of the focal
surfaces. The lines of our congruence (41) are tangent to the curves
$R^2=const$ on the focal surface with the radius-vector
$\stackrel{\rightarrow }{r}$
and can be represented parametrically as follows:
$$
\stackrel{\rightarrow }{y}=\stackrel{\rightarrow }{r}+
t\partial _1\stackrel{\rightarrow }{r},
$$
so that equations (41) can be obtained by excluding parameter $t$.
The Laplace transformation (in the direction $R^1$) of congruence (41) is
a congruence, formed by the tangents to the curves $R^2=const$
on the second focal surface $\stackrel{\rightarrow }{R}$:
$$
\stackrel{\rightarrow }{y}=\stackrel{\rightarrow }{R}+
t\partial _1\stackrel{\rightarrow }{R},
$$
or, in the components,
$$
\begin{array}{c}
y^0=\lambda^2+t\partial_1\lambda^2, \\
\ \\
y^1=\lambda^2u^1-f^1+t(\partial_1\lambda^2u^1+(\lambda^2-\lambda^1)
\partial_1u^1), \\
\ \\
y^2=\lambda^2u^2-f^2+t(\partial_1\lambda^2u^2+(\lambda^2-\lambda^1)
\partial_1u^2).
\end{array}
$$
Excluding $t$, we can rewrite these equations in the form
$$
\begin{array}{c}
y^1=U^1y^0-F^1, \\
y^2=U^2y^0-F^2,
\end{array}
$$
where
$$
\begin{array}{c}
U^1=u^1+\frac{(\lambda ^2-\lambda ^1)\partial _1u^1}{\partial _1\lambda ^2}
   =u^1-\frac{\partial _1u^1}{b},\\
\ \\
U^2=u^2+\frac{(\lambda ^2-\lambda ^1)\partial _1u^2}{\partial _1\lambda ^2}
   =u^2-\frac{\partial _1u^2}{b},\\
\ \\
F^1=f^1+\frac{\lambda ^2(\lambda ^2-\lambda ^1)\partial _1u^1}
    {\partial _1\lambda ^2}
   =f^1-\frac{\lambda ^2\partial _1u^1}{b},\\
\ \\
F^2=f^2+\frac{\lambda ^2(\lambda ^2-\lambda ^1)\partial _1u^2}
    {\partial _1\lambda ^2}
   =f^2-\frac{\lambda ^2\partial _1u^2}{b},
\end{array}
$$
(we recall that $a=\frac{\partial_2\lambda^1}{\lambda^2-\lambda^1}, ~~
b=\frac{\partial_1\lambda^2}{\lambda^1-\lambda^2}$).
The system of conservation laws
$$
\begin{array}{c}
U_t^1=F^1_x,\\
U_t^2=F^2_x,\\
\end{array}
\eqno(42)
$$
is called the Laplace transform of system (40). In Riemann invariants
the transformed system assumes the form
$$
\begin{array}{c}
R_t^1=\Lambda^1(R)R^1_x,\\
R_t^2=\Lambda^2(R)R^2_x,\\
\end{array}
$$
(we point out that the transformed system (42) has the same Riemann
invariants as (40)), where the new characteristic velocities are given by formula (5):
$$
\begin{array}{c}
\Lambda^1=\lambda^2,\\
\ \\
\Lambda^2=\lambda^2-\frac{b\partial_2\lambda^2}
{\partial_2b-ab}=
\lambda^2-\frac{1}
{\frac{\partial_1\partial_2\lambda^2}{\partial_1\lambda^2\partial_2\lambda^2}+
\frac{1}{\lambda^1-\lambda^2}}.
\end{array}
$$
We emphasize, that the Laplace transformation of the characteristic velocities
does not depend on the particular conservative representation (40) of the given
system (1).

\section{Acknowledgements}

I would like to thank the participants of the seminar
"Geometry and Mathematical Physics" in the Moscow State University
for fruitfull discussions and  N.~Kamran for sending me the reprints of
\cite{Kamran}, \cite{Kamran1}. I also take an opportunity to thank
Y. Nutku for the invitation to the T\"{U}BITAK-Marmara Research Center
where a part of this work has been completed.

This research was partially supported by RBFR-DFG grant N 96-01-00050
and the Aleksander von Humboldt Foundation.

\end{document}